\documentclass[12pt]{article}
\usepackage{amsmath} 
\usepackage{graphicx}
\usepackage{xspace}
\usepackage{hyperref}
\usepackage{float}
\usepackage{braket}
\usepackage{wrapfig}

\newcommand\pubnumber{WSU--HEP--XXYY}
\newcommand\pubdate{\today}

\newcommand*\oline[1]{%
  \vbox{%
    \hrule height .5pt
    \kern0.2ex
    \hbox{%
      \kern-0.1em
      \ifmmode#1\else\ensuremath{#1}\fi
      \kern-0.4em
    }
  }
}

\def\manc{School of Physics and Astronomy\\
The University of Manchester,\\
Oxford Road, Manchester, M13 9PL, United Kingdom
}
\def\support{\footnote{on behalf of the LHCb collaboration.}}

\def\Title#1{\begin{center} {\Large #1 } \end{center}}
\def\Author#1{\begin{center}{ \sc #1} \end{center}}
\def\Address#1{\begin{center}{ \it #1} \end{center}}

\newcommand\pubblock{\rightline{\begin{tabular}{l} \pubnumber\\
         \pubdate  \end{tabular}}}
\newenvironment{Abstract}{\begin{quotation}  }{\end{quotation}}
\newenvironment{Presented}{\begin{quotation} \begin{center} 
             PRESENTED AT\end{center}\bigskip 
      \begin{center}\begin{large}}{\end{large}\end{center} \end{quotation}}

\def\beq{\begin{equation}}
\def\eeq#1{\label{#1}\end{equation}}
\def\eeqn{\end{equation}}

\def\beqa{\begin{eqnarray}}
\def\eeqa#1{\label{#1}\end{eqnarray}}
\def\eeqan{\end{eqnarray}}

\let\bar=\overbar

\def\Dslash{\not{\hbox{\kern-4pt $D$}}}
\def\dslash{\not{\hbox{\kern-2pt $\del$}}}

\def\msb{{\bar{\ssstyle M \kern -1pt S}}}

\def\Dbar    {{\kern 0.2em\overline{\kern -0.2em D}{}}\xspace}
\def\CP                {{\ensuremath{C\!P}}\xspace}
\def\Dz      {{\ensuremath{D^0}}\xspace}
\def\Dzb     {{\ensuremath{\Dbar^0}}\xspace}
\def\Dstar   {{\ensuremath{D^*}}\xspace}

\begin{document}
\begin{titlepage}
\pubblock

\vfill
\Title{Measurements of mixing and indirect \CP violation}
\vfill
\Author{Stefanie Reichert\support}
\Address{\manc}
\vfill
\begin{Abstract}
LHCb has collected the world's largest sample of charmed hadrons. This sample is used to measure the mixing parameters in the \Dz-\Dzb system and to search for indirect \CP violation. This contribution focuses on measurements of $A_{\Gamma}$ with \Dstar and semileptonic $B$ decays and on mixing measurements and a search for \CP violation in $D\to K\pi$ decays.
\end{Abstract}
\vfill
\begin{Presented}
The 7th International Workshop on Charm Physics (CHARM 2015)\\
Detroit, MI, 18-22 May, 2015
\end{Presented}
\vfill
\end{titlepage}
\def\thefootnote{\fnsymbol{footnote}}
\setcounter{footnote}{0}

\section{Introduction}

In the Standard Model, \Dz-\Dzb mixing is suppressed due to a cancellation via the GIM mechanism \cite{GIM}\index{GIM} and CKM suppression \cite{Cabibbo,CKM}\index{Cabbibo, CKM} and therefore measurements of the mixing parameters $x$ and $y$ are experimentally challenging. Although traditionally the Standard Model predicts \CP asymmetries in \Dz-\Dzb to be $\leq 10^{-5}$ \cite{Cicerone}\index{Cicerone}, more recent theoretical calculations \cite{Lenz}\index{Lenz} find sizable effects on this limit due to the corrections from the leading Heavy Quark Expansion \cite{ManoharWise}\index{ManoharWise} contributions. Evidence for New Physics could be found if discrepancies between experimental observations and the predicted level of indirect \CP violation in the Standard Model occur.\\

In this proceedings, LHCb's latest measurements of $A_{\Gamma}$ and of the time-dependent ratio of $\Dz \to K^- \pi^+$ (right-sign) to $D^0 \to K^+ \pi^-$ (wrong-sign) decays will be reported. Charge conjugation is implied throughout this report.

\section{Theory}

In the charm sector, mixing occurs if the mass and width difference between the two mass eigenstates $\Ket{D_1}$ and $\Ket{D_2}$ with masses $m_{1,2}$ and widths $\Gamma_{1,2}$ is non zero. The mixing parameters $x$ and $y$ are defined as

\begin{align}
x &= \frac{2(m_1-m_2)}{\Gamma_1 + \Gamma_2}, \\
y &= \frac{\Gamma_1 - \Gamma_2} {\Gamma_1 + \Gamma_2}.
\end{align}

The mass eigenstates are a superposition of the flavour eigenstates $\Ket{\Dz}$ and ${\kern 0.2em\oline{\kern -0.2em \Ket{\Dz}}}$

\begin{align}
\Ket{D_{1,2}} &=  p \Ket{D^0} \pm q {\kern 0.4em\oline{\kern -0.2em \Ket{\Dz}}}
\end{align}

with complex coefficients $q,p$ satisfying the normalisation condition $|q|^2 + |p|^2 = 1$. One distinguishes between direct and indirect \CP violation. Direct \CP violation or \CP violation in decay is present if the asymmetry

\begin{align}
A_d \equiv \frac{|A_f|^2 - |\bar{A}_{\bar{f}}|^2}{|A_f|^2 + |\bar{A}_{\bar{f}}|^2},
\end{align}

is different from zero.

The amplitude of a \Dz into a final state $f$ is denoted by $A_f$ and the decay of a \Dzb into the charge-conjugated decay is $\bar{A}_{\bar{f}}$. Indirect \CP violation refers to \CP violation and \CP violation in interference. The former is defined through the asymmetry

\begin{align}
A_m \equiv \frac{|q/p|^2 - |p/q|^2}{|q/p|^2 + |p/q|^2},
\end{align}

and the interference between mixing and decay is present if the phase 

\begin{align}
\phi \equiv arg(\frac{q\bar{A}_{\bar{f}}}{pA_f})
\end{align}

is non-zero.

\section{The LHCb experiment}

The analysed data are recorded by the LHCb experiment at the Large Hadron Collider (LHC) \cite{LHC_main, LHC_general, LHC_injector}\index{LHC_main, LHC_general, LHC_injector} at CERN. Pure proton beams are produced by stripping off  the electron of hydrogen atoms and the protons are then initially accelerated by the linear collider. Successively, the protons are accelerated further in the Booster, the Proton Synchrotron and the Super Proton Synchrotron before being injected into the LHC and reaching their final collision energy.\\
 
\begin{figure}[H]
\centering
\includegraphics[width=.75\textwidth]{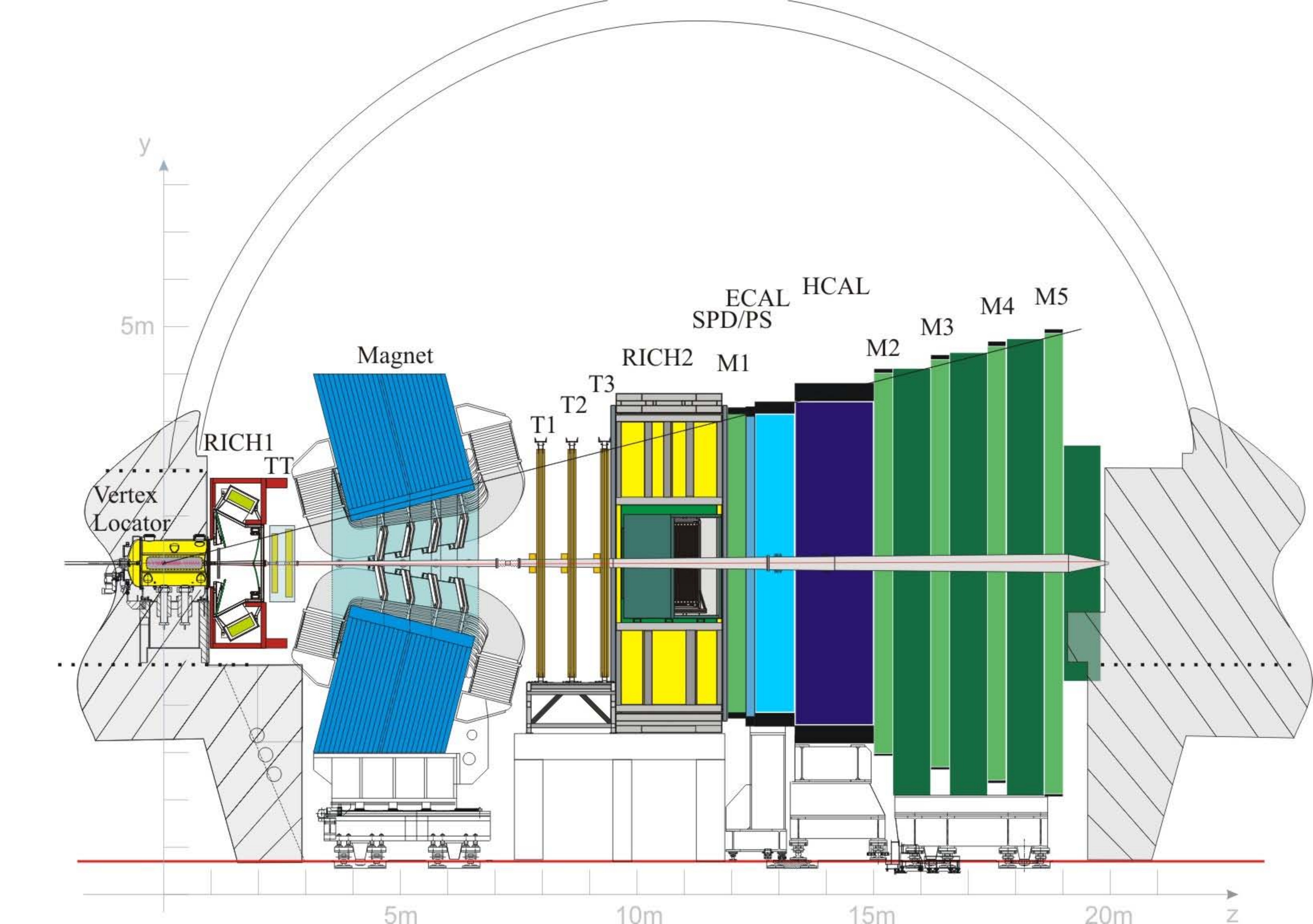}
 \caption{The LHCb detector. The figure is taken from Ref.~\cite{LHCb}\index{LHCb_detector}.}
  \label{fig:LHCb}
 \end{figure}
 
The description of the LHCb detector is based on Ref. \cite{LHCb}\index{LHCb}. The LHCb detector is a single-arm spectrometer illustrated in Fig. \ref{fig:LHCb} where the collision point is chosen as the origin of a right-handed coordinate system depicted in Fig. \ref{fig:LHCb}. With a cross section of $10\%$ of all visible events inside the LHCb detector acceptance for producing charm quarks \cite{xsec}\index{xsec}, the LHCb detector is very well suited for studies in the charm system. The beam pipe is enclosed by the Vertex Locator (VELO) aligned such that the collision point of the protons is located in the centre of the $x$-$y$ plane of the VELO. Built of silicon strip sensors, the VELO provides measurements of track and vertex coordinates with high precision. Apart from the VELO, the tracking system consists of the Tracker Turicensis (TT), and three tracking stations (T1-T3) subdivided into the Inner and Outer Trackers, (IT) and (OT). The TT as well as the IT are composed of silicon microstrip sensors whereas the OT employs straw tubes. Embedded in a $4 \, \mathrm{Tm}$ dipole magnet, this system provides both track coordinate and momentum measurements. Approximately a third of all $K^0_S$ mesons decay inside the VELO acceptance - so-called long tracks (L) - and exhibit a better momentum resolution than so-called downstream tracks (D) of $K^0_S$ mesons decaying outside the VELO acceptance. A system of Ring Imaging Cherenkov Detectors (RICH) is used to obtain excellent separation between kaons and pions. The $\mathrm{C_4F_{10}}$ and aerogel radiators of RICH1 ensure particle identification for low-momentum charged particles whereas the $\mathrm{CF_4}$ radiators of  RICH2 show a better performance for the high-momentum range. The shashlik calorimeter system composed of scintillating tiles and lead absorbers in the Electromagnetic Calorimeter (ECAL) and iron absorbers in the Hadronic Calorimeters (HCAL),  respectively, provide identification and energy measurements of electrons, photons and hadrons. To identify electrons in the trigger, a Scintillator Pad Detector (SPD) and a Preshower Detector (PS) are installed in front of the ECAL. The muon system (M1-M5) uses mostly multi wire proportional chambers filled with a gas mixture of $\mathrm{Ar:C0_2:CF_4}$ to measure the spatial coordinates of muon tracks. In the the inner region of M1, triple-GEM detectors are installed instead of multi wire proportional chambers. Iron blocks between stations M2 to M5 serve as absorbers.\\
 
$\Dz \to h^+ h^-$ with $h = (\pi, K)$ decays are classified as either prompt or semileptonic decays. The term prompt refers to a \Dz candidates originating from a $\Dstar^+ \to \Dz \pi^+$ decay where the $\Dstar^+$ was directly produced in the $pp$ collision. If the \Dz candidates originates from a semileptonic $B$ decay where the $B$ meson was produced in the $pp$ collision, the \Dz is a semileptonic candidate. A measure to distinguish between prompt and semileptonic decays is the impact parameter (IP). The IP is defined as the distance of closest approach of the reconstructed trajectory to the $pp$ collision vertex. For semileptonic decays, the impact parameter is large whereas for prompt decays, it is close to zero.

\section{$\mathbf{A_{\Gamma}}$ measurements}
 
The asymmetry of the inverse effective lifetimes of \Dz and \Dzb, $\hat{\Gamma}_{\Dz}$ and $\hat{\Gamma}_{\Dzb}$, into \CP-even final states, $A_{\Gamma}$ is defined as 
 
 \begin{align}
 A_{\Gamma} = \frac{\hat{\Gamma}_{\Dz} - \hat{\Gamma}_{\Dzb}}{\hat{\Gamma}_{\Dz} + \hat{\Gamma}_{\Dzb}} \approx \left ( \frac{A_d+A_m}{2}\right ) y\cos(\varphi) - x \sin(\varphi), \quad \varphi = arg(q,p).
 \end{align}
 
A measurement of $A_{\Gamma}$ is an almost pure measurement of indirect \CP violation since measurements have shown that $A_d$ is small compared to the precision of $A_{\Gamma}$ \cite{HFAG}\index{HFAG}. In the Standard Model, the \CP-violating phase is independent of the final state thus $ A_{\Gamma}(\pi^+\pi^-) = A_{\Gamma}(K^+K^-) = A_{\Gamma}$ is expected to hold.

 \subsection{$\mathbf{A_{\Gamma}}$ from prompt decays}
 
The measurement of $A_{\Gamma}(\pi^+\pi^-)$ and $A_{\Gamma}(K^+K^-)$~\cite{PromptAG}\index{PromptAG} is based on the dataset recorded in 2011 with the LHCb detector. The data was taken at a centre-of-mass energy of $\sqrt{s} = 7 \rm \, TeV$ and corresponds to an integrated luminosity of $1\rm fb^{-1}$.\\

\begin{wrapfigure}[16]{r}{0.55\textwidth}
\vspace{-42pt}
  \begin{center}
    \includegraphics[width=0.5\textwidth]{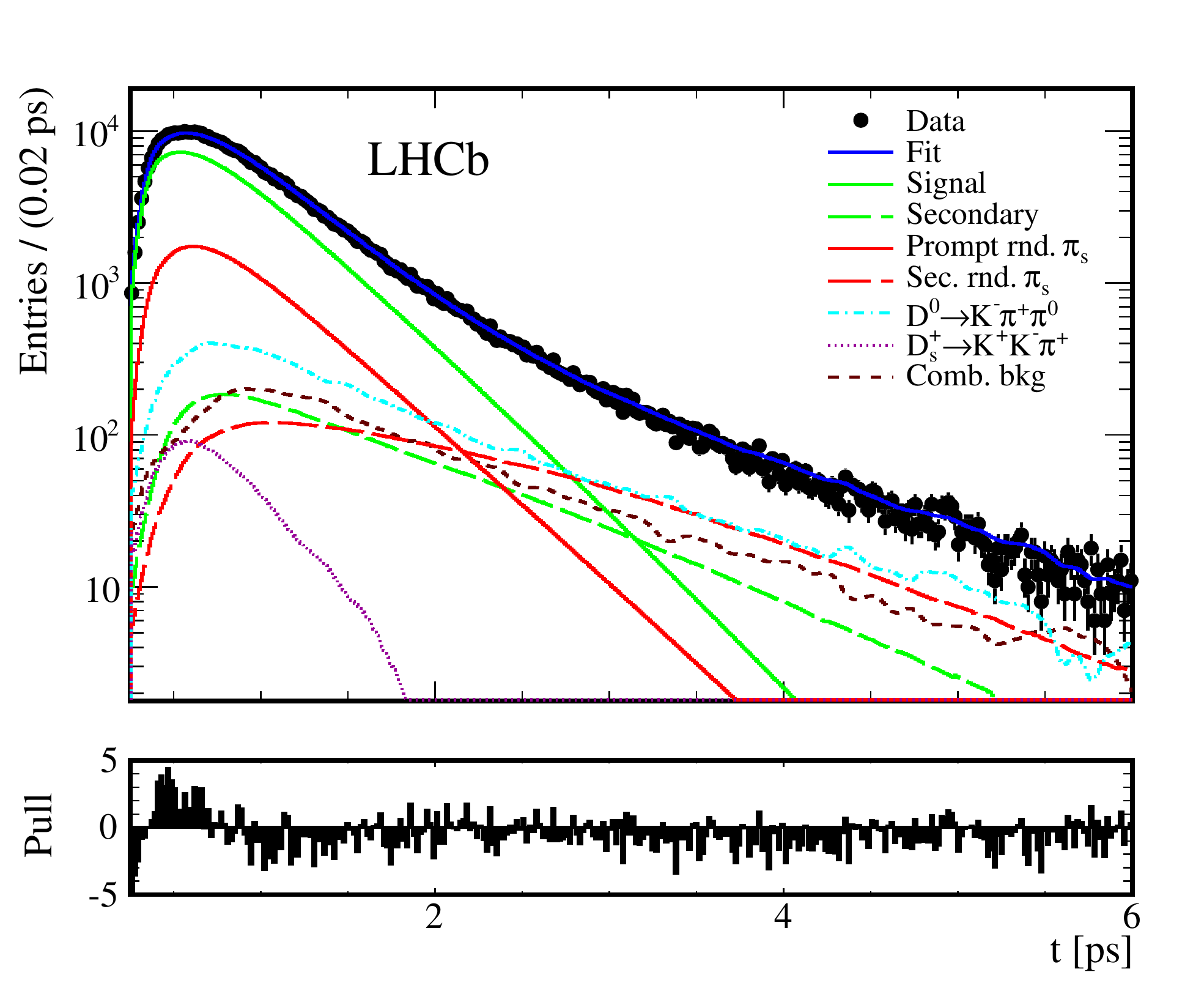}
  \end{center}
  \vspace{-21pt}
 \caption{Fit projection on decay time to $\Dzb \to K^- K^+$ and corresponding pull plot for candidates for a subset of the data~\cite{PromptAG}.} 
   \label{fig:promptAGlifetime}
\end{wrapfigure}

The \Dz flavour is determined from the charge of the $\pi^+$ from the $\Dstar^+$ decay. The measurement relies on $3.1\rm M$ prompt $D^0 \to K^+ K^-$ signal candidates with a purity of $93\%$ and on $1.0\rm M$ $D^0 \to \pi^+ \pi^-$ signal candidates with a purity of $91\%$. The determination of the effective lifetimes $\hat{\Gamma}_{\Dz}$ and $\hat{\Gamma}_{\Dzb}$ is carried out in two steps: a multivariate unbinned maximum likelihood fit to $m(h^+h^-)$ with $h = K, \pi$ and $\Delta m \equiv m(\Dstar) - m(\Dz)$ is performed to extract the signal yields. In a second step, the effective lifetime is extracted from a multivariate unbinned maximum likelihood fit to \Dz decay time and to $\ln(\chi^2_{\rm IP})$. The observable $\chi^2_{\rm IP}$ is defined as the difference in $\chi^2$ of a given $pp$ collision vertex reconstructed with and without the considered track. Fitting in decay time and $\ln(\chi^2_{\rm IP})$ allows the separation of prompt and semileptonic components. Figure~\ref{fig:promptAGlifetime} illustrates the fit projection on decay time for $\Dzb \to K^- K^+$ decays.

The systematic uncertainties arise from partially reconstructed and other backgrounds, charm decays from semileptonic decays and mainly from a per-event data-driven acceptance function. The results are $A_{\Gamma}(K^+K^-) = (-0.035 \pm 0.062 \pm 0.012)\%$ and $A_{\Gamma}(\pi^+\pi^-) = (0.033 \pm 0.016 \pm 0.014)\%$ with statistical and systematic uncertainties. The measurements are in agreement with the no indirect \CP violation hypothesis. 
 
 \subsection{$\mathbf{A_{\Gamma}}$ from semileptonic decays}
 
A measurement of $A_{\Gamma}(\pi^+\pi^-)$ and $A_{\Gamma}(K^+K^-)$~\cite{SLAG}\index{SLAG} is performed on the 2011 and 2012 dataset. In 2011, $1\rm fb^{-1}$ was recorded at $\sqrt{s} = 7 \rm \, TeV$ whereas the centre-of-mass energy in 2012 was increased to $\sqrt{s} = 8 \rm \, TeV$ and $2\rm fb^{-1}$ of data were taken.\\

\begin{wrapfigure}[25]{r}{0.55\textwidth}
\begin{minipage}[t]{\linewidth}
\centering
\vspace{-23pt}
\includegraphics[width=0.87\linewidth]{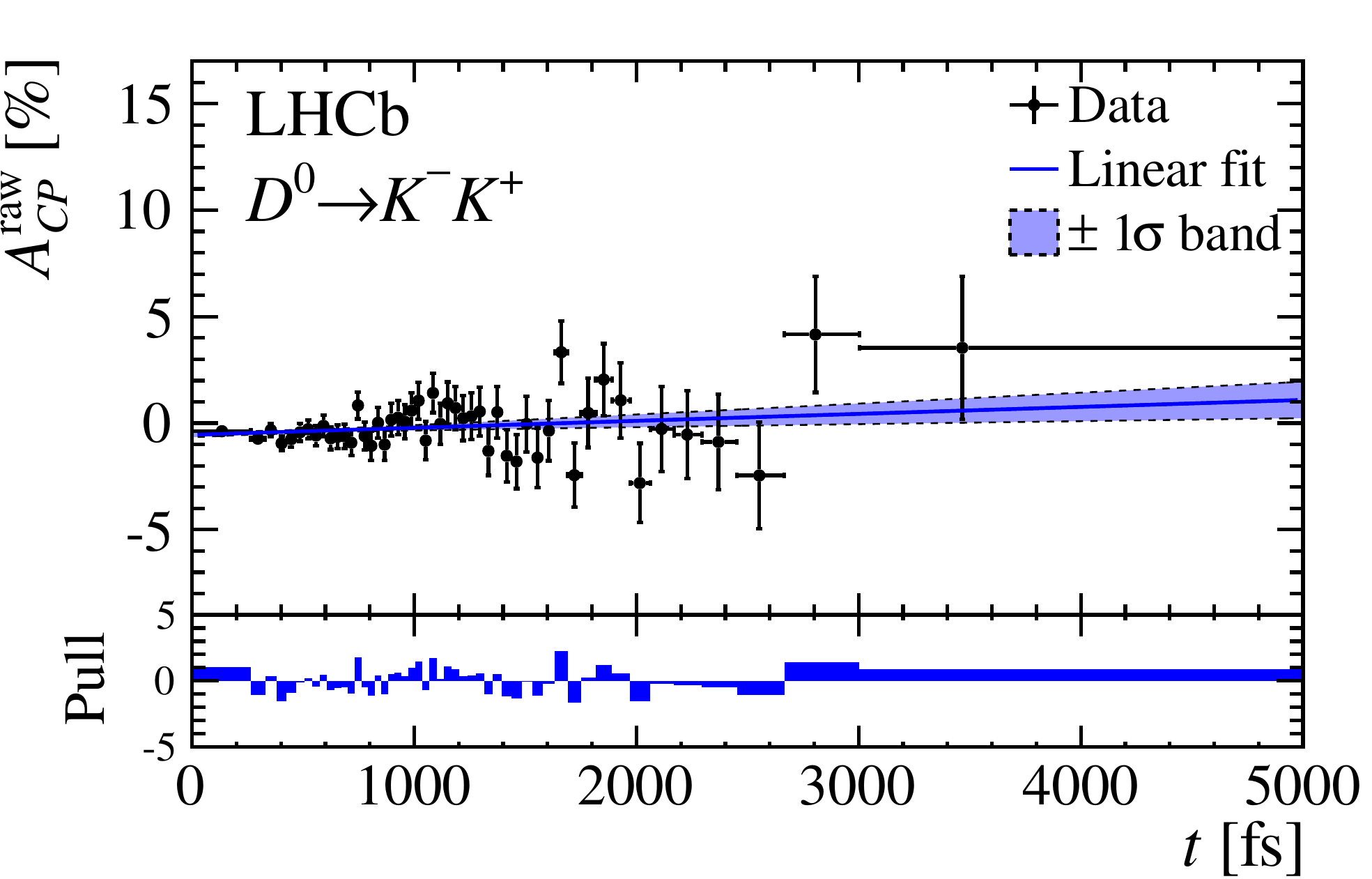}
\vspace{-10pt}
\caption{Raw \CP asymmetry as a function of \Dz decay time for $D^0 \to K^+ K^-$ candidates~\cite{SLAG}.}
 \label{fig:SLAG_KK}
\end{minipage}
\begin{minipage}[t]{\linewidth}
\centering
\includegraphics[width=0.87\linewidth]{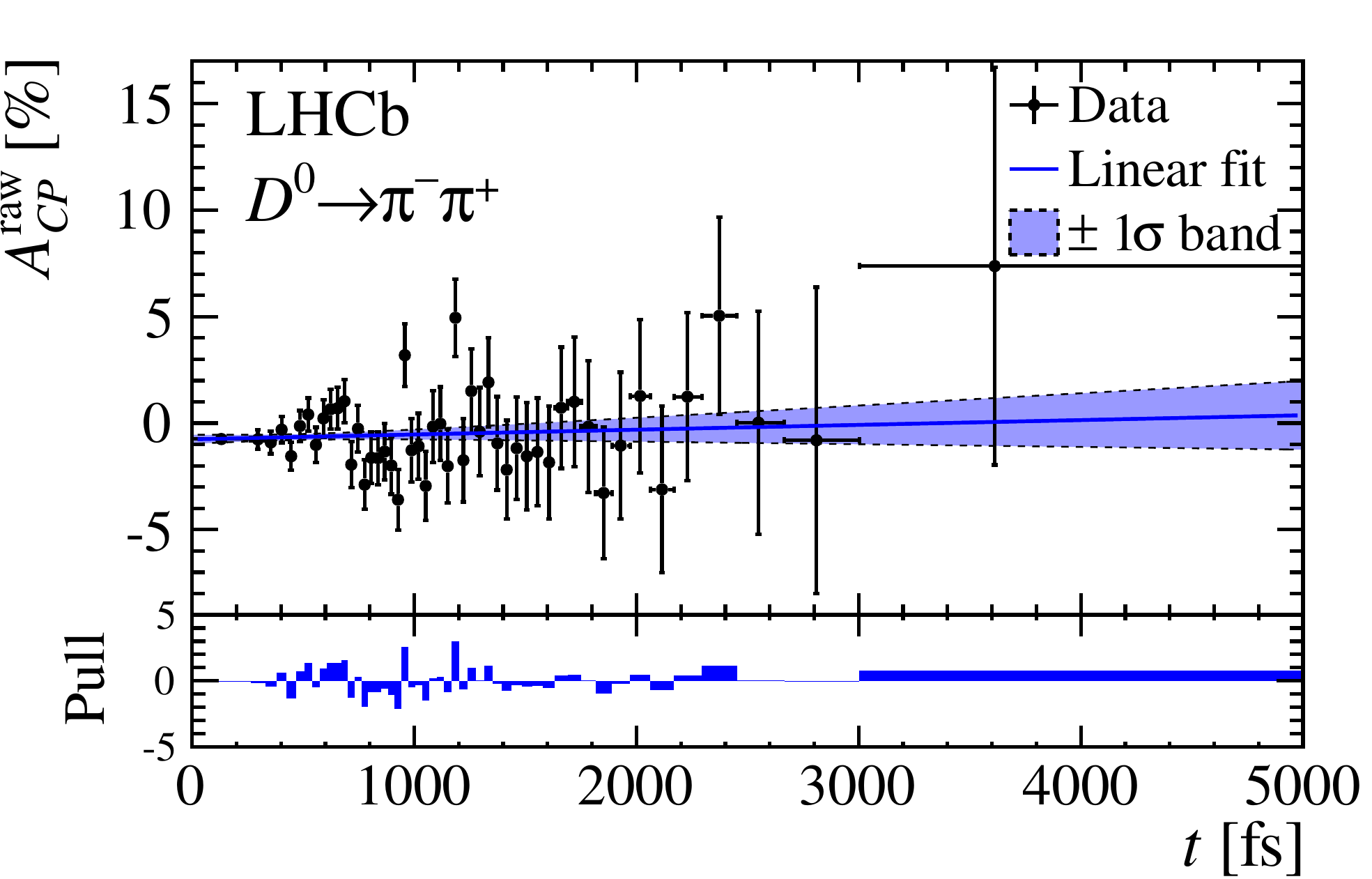}
  \vspace{-10pt}
\caption{Raw \CP asymmetry as a function of \Dz decay time for $D^0 \to \pi^+ \pi^-$ candidates~\cite{SLAG}.}
\label{fig:SLAG_pipi}
\end{minipage}
\end{wrapfigure}

In this analysis, \Dz decays are required to originate from semileptonic $B^- \, (\bar{B}^0) \to D^0 \mu^- \bar{\nu}_{\mu} X$ decays and the \Dz flavour is determined from the muon charge. The dataset contains $2.3\rm M$ $D^0 \to K^+K^-$ and $0.8\rm M$ $D^0 \to \pi^+\pi^-$ signal candidates. $A_{\Gamma}$ is determined from a $\chi^2$ fit to the time-dependent raw \CP asymmetry $A_{\CP}^{\rm raw}(t)$

\begin{align}
A_{\CP}^{\rm raw}(t) \approx A_d - A_{\Gamma} \frac{t}{\tau}.
\end{align}

The raw \CP asymmetry is determined from simultaneous fits to $m(h^+h^-)$ with $h=K, \pi$ in 50 bins of \Dz decay time to \Dz and \Dzb samples. The time-dependent $A_{\CP}^{\rm raw}(t)$ is shown in Figs.~\ref{fig:SLAG_KK} and~\ref{fig:SLAG_pipi}. The systematic uncertainties arise from the \Dz mass fit model and the decay time resolution, $B^0-\bar{B}^0$ mixing, detection and production asymmetries, and the time-dependent efficiency. The main contributions to the systematic uncertainty result from the mistag probability and the mistag asymmetry. The mistag probability is defined as the probability of the muon to have the wrong charge with respect to the true \Dz flavour.\\

\begin{wrapfigure}[12]{r}{0.55\textwidth}
  \begin{center}
  \vspace{-18pt}
    \includegraphics[width=0.5\textwidth]{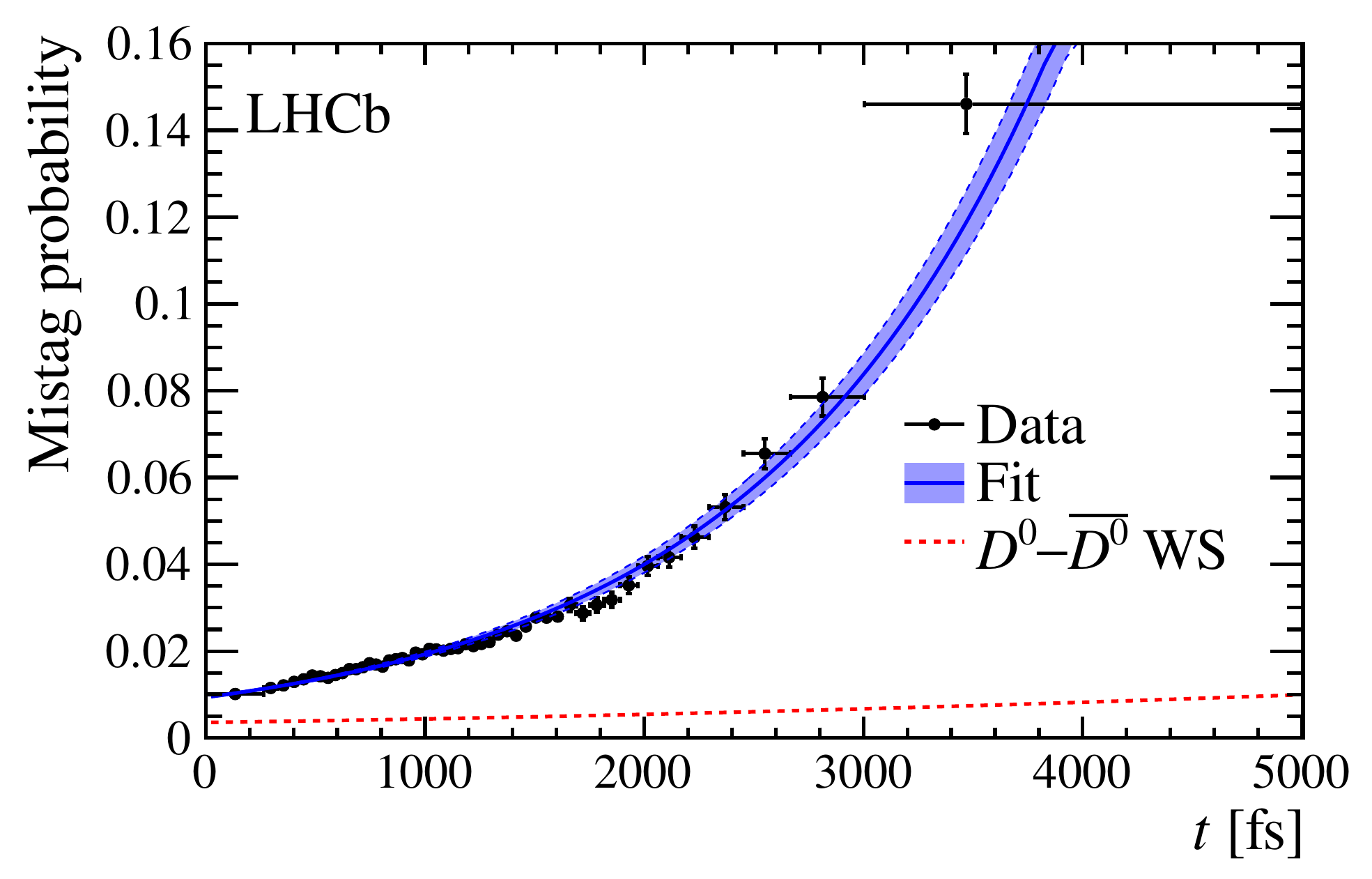}
  \end{center}
  \vspace{-19pt}
 \caption{Mistag probability overlaid with an exponential fit~\cite{SLAG}.} 
   \label{fig:SLAG_mistag}
\end{wrapfigure}

A control sample of $D^0 \to K^-\pi^+$ decays is used to measure the mistag probability depicted in Fig.~\ref{fig:SLAG_mistag}. Since the mistag probability can differ for positive and negative muons, the slope of this mistag asymmetry is calculated on the control sample. The slope of the mistag asymmetry is found to be non time-dependent. The statistical uncertainty on the slope is included in the systematic uncertainty and is the dominant contribution.\\

$A_{\Gamma}$ is measured to be $A_{\Gamma}(K^+K^-) = (-0.134 \pm 0.077 \,^{+0.026}_{-0.034})\%$ and $A_{\Gamma}(\pi^+\pi^-) = (-0.092 \pm 0.145 \,^{+0.025}_{-0.033})\%$ where the uncertainties are statistical and systematic, respectively. This measurement is in agreement with the conservation of \CP symmetry.

 \section{Mixing measurement and search for \CP violation in $\mathbf{D\to K\pi}$ decays}
  
Mixing can be measured from the time-dependent ratio of $\Dz \to K^- \pi^+$ (right-sign) to $D^0 \to K^+ \pi^-$ (wrong-sign) decays. The right-sign decay $\Dz \to K^- \pi^+$ is a Cabibbo-favoured decay whereas the decay $\Dz \to \Dzb \to K^- \pi^+$ is doubly-Cabibbo suppressed. For the wrong-sign decays, $D^0 \to K^+ \pi^-$ is doubly-Cabibbo suppressed and $\Dz \to \Dzb \to K^+ \pi^-$ is a Cabbibo-favoured mode. In the limit of small mixing and for negligible \CP violation, the time-dependent ratio of right-sign to wrong-sign decays $R(t)$~\cite{WS2011}\index{WS2011} is given by

\begin{align}
R(t) \approx R_D + \sqrt{R_D} \, y' \frac{t}{\tau} + \frac{x'^{\,2} + y'^{\,2}}{4} \left (  \frac{t}{\tau}  \right )^2.
\end{align}

The mixing parameters are rotated by the strong phase-difference $\delta$ between the doubly-Cabibbo suppressed and Cabibbo-favoured amplitudes\\
 $\mathcal{A}(\Dz \to K^+\pi^-)/\mathcal{A}(\Dzb \to K^+\pi^-) =  -\sqrt{R_D} \, e^{-i\delta}$ and are given by

\begin{align}
x' &\equiv x\cos \delta + y \sin \delta,\\
y' &\equiv y \cos \delta - x \sin \delta. 
\end{align}

The analyses presented in the following allow to measure $x'^{\, 2}$, $y'$. The flavour of the prompt $D \to K\pi$ decay is determined by the charge of the pion from the $\Dstar^+$ decay.

 \subsection{Mixing in $\mathbf{D \to K\pi}$ on 2011 data}
 
On the 2011 dataset corresponding to an integrated luminosity of $1\rm fb^{-1}$, $8\rm M$ right-sign and $36\rm k$ wrong-sign decays are found~\cite{WS2011}. The signal yields are extracted in 13 bins of \Dz decay time. From a binned $\chi^2$ fit to the time-dependent ratio of right-sign to wrong-sign decays illustrated in Fig.~\ref{fig:WS_2011}, $x'^{\, 2}$ and $y'$ are extracted.\\
 
 \begin{wrapfigure}[14]{r}{0.55\textwidth}
  \begin{center}
    \vspace{-35pt}
    \includegraphics[width=0.47\textwidth]{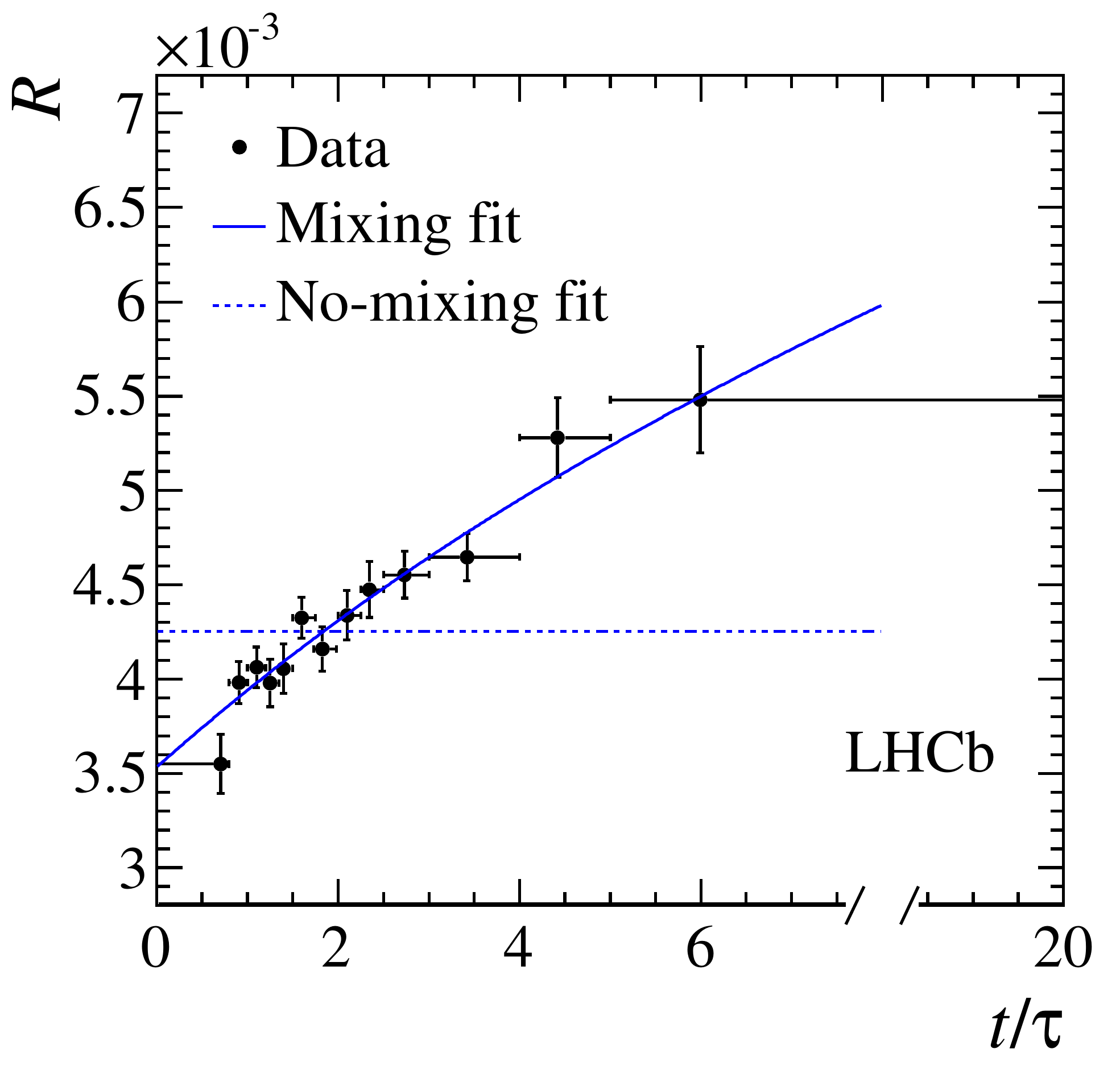}
  \end{center}
      \vspace{-30pt}
 \caption{Ratio $R$ as a function of \Dz decay time~\cite{WS2011}.} 
   \label{fig:WS_2011}
\end{wrapfigure}

In the ratio, most systematic uncertainties are cancelling. Non-cancelling contributions arise from misidentified or partially reconstructed \Dz mesons and from charm decays originating from semileptonic decays.\\

The results are $R_D = (3.52 \pm 0.15)\times 10^{-3}$, $x'^{\,2} = (-0.9 \pm 1.3)\times 10^{-4}$ and $y' = (7.2 \pm 2.4 )\times 10^{-3}$ where the uncertainties include both statistical and systematic uncertainties. This measurement excludes the no-mixing hypothesis at $9.1\sigma$.

 \subsection{Mixing and \CP violation in $\mathbf{D \to K\pi}$ on 2011 and 2012 data}

 The analysis was performed on the 2011 and 2012 datasets corresponding to an integrated luminosity of $3\rm fb^{-1}$. The mixing parameters $x'^{\, 2}$ and $y'$ are extracted from $53\rm M$ right-sign and $230\rm k$ wrong-sign signal candidates~\cite{WS2012}\index{WS2012}. To search for direct and indirect \CP violation, the data is split into \Dz and \Dzb samples. For each sample, the time-dependent ratios $R^{\pm}(t)$ for \Dz and \Dzb, respectively, are measured and fitted for parameters $R_D^{\pm}, x'^{\, 2\pm}$ and $y'^{\pm}$. The mixing parameters are defined as
 
\begin{align}
x'^{\pm} &\equiv \left \vert q/p  \right \vert^{\pm 1} x' \cos \phi \pm y' \sin \phi \\
y'^{\pm} &\equiv \left \vert q/p  \right  \vert^{\pm 1}  y' \cos \phi \mp x' \sin \phi. 
\end{align}
 
 \begin{wrapfigure}[18]{r}{0.55\textwidth}
  \begin{center}
  \vspace{-18pt}
    \includegraphics[width=0.47\textwidth]{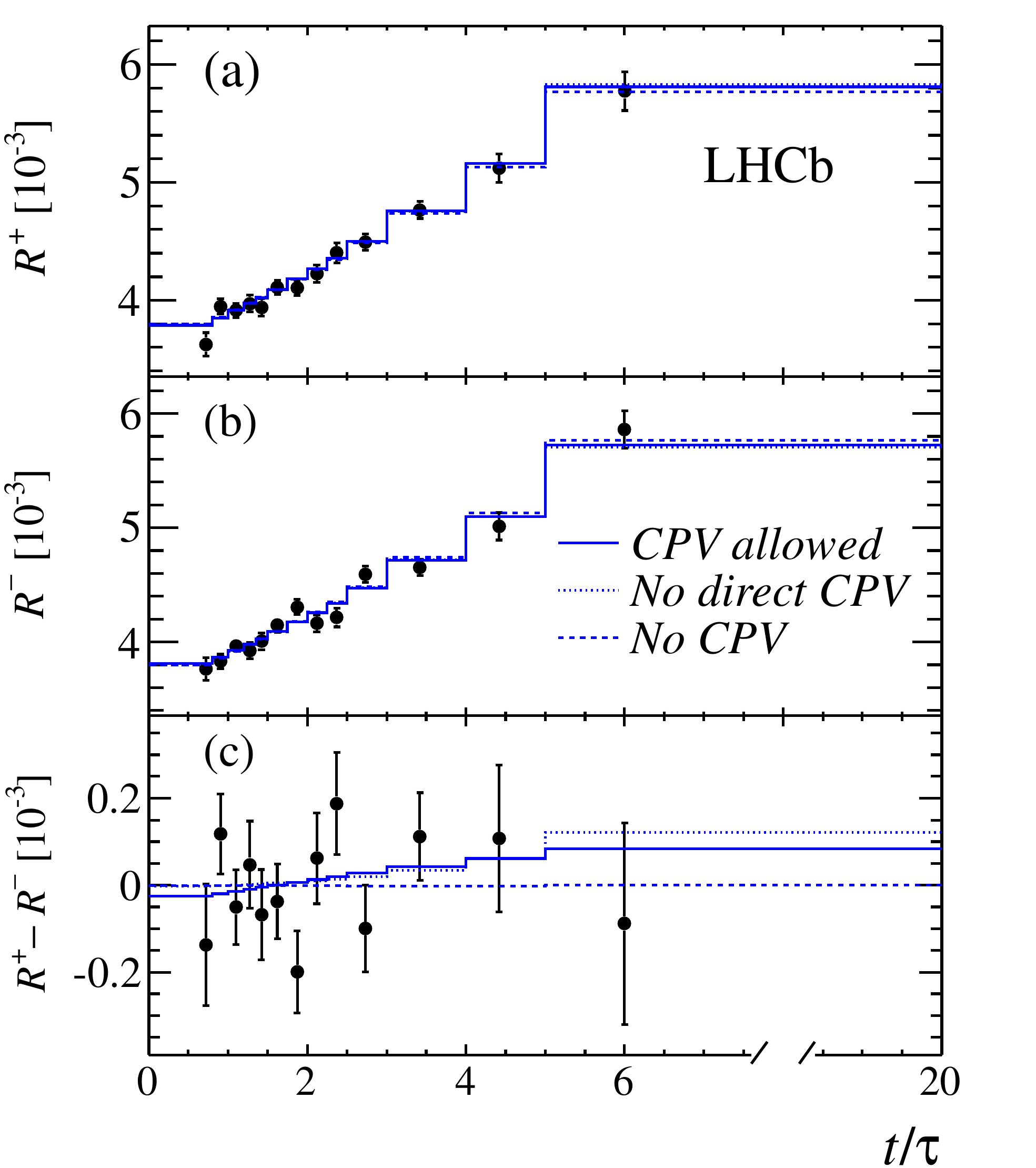}
  \end{center}
    \vspace{-23pt}
 \caption{Ratios $R+$ (top), $R^-$ (centre) and their difference (bottom) as a function of \Dz decay time~\cite{WS2012}.} 
   \label{fig:WS_2011}
\end{wrapfigure}

The fit allows three different scenarios: \CP violation allowed, no direct \CP violation and no \CP violation. The asymmetry $A_D \equiv (R^+ - R^-/(R^++R^-))$ is a measure for direct \CP violation whereas a difference in $(x'^{\,2 +}, y'^+)$ and $(x'^{\,2 -}, y'^-)$ implies indirect \CP violation. Assuming \CP conservation, the fit results are $R_D = (3.568 \pm 0.066)\times 10^{-3}$, $x'^{\,2} = (5.5 \pm 4.9)\times 10^{-5}$ and $y' = (4.8 \pm 1.0 )\times 10^{-3}$.\\

Allowing for \CP violation, the direct \CP asymmetry is $A_D = (-0.7 \pm1.9)\%$ and at $68.3\%$ C. L., $0.75 < \left \vert q/p \right \vert < 1.24$ wherefore no evidence for \CP violation is found.
 
 \section{Conclusion}
 
 Searches for \CP violation in charm provide probes of new physics and LHCb has contributed significantly to the understanding of mixing and \CP violation. The world average of $A_{\Gamma}$ is dominated by the LHCb measurements with a combined precision of $4\times 10^{-4}$. The measurements of $A_{\Gamma}$ and the search for \CP violation in $D\to K\pi$ decays are compatible with the conservarion of \CP symmetry. The no mixing hypothesis was excluded by the 2011 $D \to K\pi$ analysis by $9.1\sigma$.


\end{document}